\documentclass[preprint]{aastex}

\received{}
\revised{}
\accepted{}
\ccc{}
\cpright{}{}

\slugcomment{AJ, in press (Sep. 2001)}
\shorttitle{Fe Abundance of M71}
\shortauthors{Ram\'{\i}rez \etal}

\newcommand{\kms}{km~s$^{-1}$}

\newcommand{\etal}{{\it et al.\/}}
\newcommand{\teff}{$T_{eff}$}
\newcommand{\grav}{log($g$)}
\newcommand{\mtv}{$\xi$}
\newcommand{\ew}{$W_{\lambda}$}
\newcommand{\fe}{[Fe/H]}

\begin{document}

\title{Abundances in Stars from the Red Giant Branch Tip to the Near the 
Main Sequence Turn Off in M71: II. Iron Abundance
\altaffilmark{1}}

\author{Solange V. Ram\'{\i}rez \altaffilmark{2}, 
Judith G. Cohen\altaffilmark{2}, 
Jeremy Buss\altaffilmark{3} and
Michael M. Briley\altaffilmark{3}}

\altaffiltext{1}{Based on observations obtained at the
W.M. Keck Observatory, which is operated jointly by the California 
Institute of Technology and the University of California}

\altaffiltext{2}{Palomar Observatory, Mail Stop 105-24,
California Institute of Technology}

\altaffiltext{3}{Department of Physics, University of Wisconsin,
Oshkosh, Wisconsin}

\begin{abstract}
We present \fe\ abundance results that involve a
sample of stars with a wide range in luminosity from luminous giants to
stars near the turnoff in a globular cluster.
Our sample of 25 stars in M71 includes 10 giant stars more luminous than the 
RHB, 3 horizontal branch stars, 9 giant stars less luminous than the RHB,
and 3 stars near the turnoff. 
We analyzed both Fe I and Fe II lines in high dispersion spectra observed 
with HIRES at the W. M. Keck Observatory.
We find that the \fe\ abundances from both Fe I and Fe II
lines agree with each other and with earlier determinations.
Also the \fe\ obtained from Fe I and Fe II lines is constant within the rather
small uncertainties for this group of stars over the full range in
\teff\ and luminosity, suggesting that NLTE effects are negligible in 
our iron abundance determination.
In this globular cluster, there is no difference among the
mean \fe\ of giant stars located at or above the RHB, RHB stars, 
giant stars located below the RHB and stars near the turnoff.
\end{abstract}


\section{INTRODUCTION}

Abundance determinations of stars in Galactic globular clusters can provide 
valuable information about important astrophysical processes such as
stellar evolution, stellar structure, Galactic chemical evolution and
the formation of the Milky Way. Surface stellar abundances of C, N, O,
and often Na, Mg, and Al are found to be variable among red giants within 
a globular cluster. 
The physical process responsible of these star-to-star element variations 
is still uncertain \citep[see ][Paper I]{coh01}.

%

Of particular importance to the present study are the results of
\citet{kin98}, who found that \fe\ among M92 subgiants
is a factor of two smaller than \fe\ from red giants in the same
cluster \citep{coh79,sne91}. If this result is not
due to systematic differences arising from the analysis procedures of the
different groups that handle the red giant branch (RGB) and the subgiant
samples,
then the results of \citet{kin98} would suggest some modification of
photospheric Fe abundances which would be quite difficult to explain.
Clearly the present day clusters stars are incapable of modifying their
own Fe abundances through nucleosynthesis. Yet evolution-driven
changes in \fe\ could be possible if the outer envelopes of the subgiants
were somehow infused with Fe-poor external material. Thus as evolution progesses up
the RGB the deeping convective envelope would dilute the photospheric \fe\
with more Fe-rich ``unmodified'' material from the interior causing an increase
in \fe\ with luminosity. We note the unlikelyhood of this scenario as
Fe, unlike the lighter elements, is not involved in normal mass transfer
processes that might ocurr in binary systems or in normal stellar winds,
and is only produced in supernovae. Moreover, it is difficult to understand
how cluster stars could be contaminated by material with {\it less} Fe,
presumably at a time after the present subgiants had formed.

In order to study the origin of the star-to-star abundance variations
and to address the issues raised by \citet{kin98}, 
we have started a program to determine chemical abundances of the nearer
galactic globular cluster stars.  In this paper, we present our results for
the iron abundance of M71, the nearest globular cluster reachable from the 
northern hemisphere. 
Similar programs are underway at ESO by \citet{cas00} and
\citet{gra01} taking advantage of the fact that the
nearest globular cluster accessible from a southern site (NGC 6397)
is 1.8 times close than M71 and has a lower reddening.
Our M71 sample includes stars over a large range in luminosity:
19 giant stars, 3 horizontal branch stars, and 3 stars near the main
sequence
turnoff, in order to study in a consistent manner red giants, horizontal
branch stars,
and stars at the main sequence turnoff.
Details on the star sample, observations, data reduction 
and determination of stellar parameters are described in 
Paper I. 
Previous high dispersion abundance analysis for M71 involve studies of 
red giants only, resulting in \fe\ of $-$0.70 \citep{coh83}, 
$-$0.80 \citep{gra86}, between $-$0.6 and $-$1.0 \citep{lee87}, 
and $-$0.79 \citep{sne94}.

\section{ANALYSIS}

We begin our analysis with Fe, as many Fe lines are identified in our
HIRES spectra over a wide  range of excitation potentials and line
strengths,
as well as over two ionization states (Fe I and Fe II).
The iron abundance analysis is done using a current version of the LTE 
spectral synthesis program MOOG \citep{sne73}. 
A line list specifying the wavelengths, excitation
potentials, gf--values, damping constants, and equivalent widths for the 
observed Fe I and Fe II lines is required. 
The provenance of the gf--values and the measurement of equivalent widths 
are discussed below. 
The damping constants for all Fe I and Fe II lines were set to twice 
that of the Uns\"{o}ld approximation for van der Waals broadening 
following \citet{hol91}. 
The use of the Blackwell approximation for the damping constants gives 
the same result within the errors, when comparing \fe\ obtained with the 
``good line set'' of Fe I (see definition of line sets on Sec. 2.2).

In addition, a model atmosphere for the effective temperature
and surface gravity appropriate for each star and a value for the 
microturbulent velocity are also required.
We use the grid of model atmospheres from \citet{kur93a} with a metallicity of
\fe\ = $-$0.5 dex, based on earlier high dispersion abundance analysis of M71
red giants \citep{coh83,gra86,lee87,sne94}. 
The final result for \fe\ is not sensitive to small changes in the metallicity of
the model atmosphere. In particular, the error introduced by using a model
with \fe\ = $-$0.5 instead of $-$0.7 dex is very small (see Tables 2 \& 3).
The effective temperatures and surface gravities are
derived from the photometry of the stars as described in Paper I. The error 
in the photometric \teff\ is 75 K for giants and 150 K for the dwarfs and 
the error in the photometric \grav\ is 0.2 dex (Paper I).
The microturbulent velocity is derived spectroscopically (see below). 
The stellar parameters are listed in Table 1.

\subsection{Transition Probabilities}

Transition probabilities for the Fe I lines were obtained from several 
laboratory experiments, including studies of Fe I absorption 
lines produced by iron vapor in a carbon tube furnace 
\citep{bla79,bla82a,bla82b,bla86} (Oxford Group),
measurement of radiative lifetimes of Fe I transitions by laser induced 
fluorescence \citep{obr91,bar91,bar94}, 
Fe I emission line spectroscopy from a low current arc \citep{may74}, 
and emission lines of Fe I from a shock tube \citep{wol71}.
We also considered solar gf--values from \citet{the89,the90} when needed.

We compare the gf--values obtained by the different experiments in an attempt
to place them onto a common scale with respect to the results from 
\citet{obr91}, who provided the longest list of gf--values. 
We considered for the comparison 
only the set of lines present in our data, 
which have a wavelength coverage roughly from 5380 \AA\ to 7900 \AA.
We found that the values of \citet{obr91} and of the Oxford Group were 
on the
same scale; the mean difference in log($gf$) between the two
experiments is $0.02 \pm 0.01$ for 21 lines in common. Similar 
results are found when considering 21 lines in common between \citet{obr91}
and  \citet{bar91} \citep[see also ][]{bar94}, where the mean difference in 
log($gf$) is $0.02 \pm 0.04 $.
Considering 34 lines in common between \citet{obr91}
and  \citet{may74}, the mean difference in log($gf$) is $0.03 \pm 0.03$.
When comparing the 14 lines in common between \citet{obr91} and 
\citet{wol71}
we found a mean difference in log($gf$) of $-0.07 \pm 0.02$.
We also compared the results from 
\citet{obr91} with solar gf--values obtained by \citet{the89,the90}. 
We found 
that the mean difference in log($gf$) is $+0.05 \pm 0.02$ when comparing 
68 lines in common between these latest works. Similar offsets are found by 
\citet{lam96}.

The gf--values for our Fe I lines were taken when possible from laboratory 
data in the following order of priority: from \citet{obr91}, from the Oxford 
Group, from \citet{may74}, and from \citet{wol71} corrected by $-0.07$ dex.
In the cases where no laboratory data were available, we used solar gf--values
from \citet{the89,the90} corrected by +0.05 dex.

Transition probabilities for the Fe II lines were taken from the solar analysis of 
\citet{bla80}, \citet{bie91}, and from the semiempirical calculations of 
\citet{kur93b}.
While restricting ourselves to lines present in our spectra, we compared the
gf--values from \citet{bla80} and \citet{kur93b} to the study of
\citet{bie91}.
There are 9 lines in common between \citet{bie91} and \citet{bla80}. The
mean difference between the two in log($gf$) is $0.14 \pm 0.02$. A similar
offset was found by \citet{lam96}. We found no significant difference between
the results of \citet{bie91} and \citet{kur93b}, since the 9 lines in common
result in a mean difference of $0.03 \pm 0.02$. 

The gf--values for our Fe II lines were taken in the following order of priority:
from \citet{bie91}, from \citet{bla80} corrected by 0.14 dex, and from 
\citet{kur93b}.

\subsection{Measurement of Equivalent Widths}

Our sample contains many stars observed in mutiple orders, with many
detectable absorption features in these high S/N spectra. For example,
in the coolest M71 star in our sample, M71-1-45, 1407 absorption lines
have been identified.
A FORTRAN code to automatically search for absorption features and 
measure their equivalent width (\ew), EWDET, was developed for this 
project.
The code is available upon request to SVR. 
EWDET determines the continuum location of the HIRES spectra by fitting
a curve to the spectra performing several iterations of point rejection
above and below sigma levels given by the user. Then, EWDET identifies
lines above the noise level defined by a factor of two of the continuum rms
dispersion. 
Each of the identified lines is fit by a Gaussian profile and then
the \ew\ are computed by the integration of the fitted Gaussian.
The error in \ew\ is computed by adding quadratically the error
at each point of integration, $\sigma_{i}$, times the step of the integration.
The error at each point of integration is given by:
$$\sigma_{i}^{2} = g(\lambda_{i})^{2} \times 
\left[ \frac{\sigma_{P}^{2}}{P^{2}} +
\frac{(\lambda_{i} - \lambda_{cen})^{2}}{\sigma^{4}} \sigma_{\lambda_{cen}}^{2}
+
\frac{(\lambda_{i} - \lambda_{cen})^{4}}{\sigma^{6}} \sigma_{\sigma}^{2} +
\sigma_{cont}^{2} \right] $$
where $g(\lambda_{i})$ is the Gaussian profile, $P$ is the peak of the
Gaussian, $\sigma_{P}$ is the error in the peak of the Gaussian, 
$\lambda_{cen}$ is the central wavelength of the Gaussian, 
$\sigma_{\lambda_{cen}}$ is the error in the central wavelength, 
$\sigma$ is the dispersion of the Gaussian, $\sigma_{\sigma}$ is
the error in the dispersion of the Gaussian, and $\sigma_{cont}$ is the error in
the continuum. The errors of the Gaussian parameters are from the
covariance matrix of the Gaussian fit. The expression for the error in \ew\
is deduced by propagating the errors of the Gaussian parameters and 
assuming that the continuum level is equal to one (see Appendix).
The fit by a Gaussian profile is reasonable even for the strongest lines
we use, as shown in Figure 1, where the observed line is plotted with a solid 
curve and its corresponding Gaussian profile fit is shown with a dashed curve.

The line list identified and measured by EWDET is then correlated to the line 
list with the atomic parameters to specifically identify the Fe I and 
Fe II lines. 
The detailed lists of \ew\ and gf--values will be given in the next paper.

The spectral resolution, $\lambda / \Delta \lambda $, of an echelle is fixed,
unlike a low incidence, angle of low order grating spectrograph where 
$\Delta \lambda $ is constant, independent of $\lambda $. 
Hence a line of constant central depth, $D$, will have an equivalent width 
proportional to $\lambda D$.
We construct the relationship between $\lambda D$ and \ew\ for the Fe I lines
to look for possible blends and for saturation effects. 
The $\lambda D$ versus \ew\ relation for three stars
are plotted in Figure 2; M71-I is one of the most luminous and coolest stars 
of our sample, M71-G53476\_4543 is a medium luminosity and temperature star, 
and M71-G53392\_4624 is one of the faintest and hottest stars of the sample. 
We fit a second order polynomial to the $\lambda D-$\ew\ relationship
for each star,  performing several iterations of point rejection above 
and below the two sigma level. The second order term is needed to account for
line saturation at large \ew\ and the rejection of points is needed to 
eliminate blended lines. The second order fit is plotted
as a solid line in Figure 2, and the points considered for such a fit are
shown in black.

For Fe I, we use two sets of lines. The first set, subsequently called 
``the weak line set'', contains those Fe I lines which
are within two sigma levels of the $\lambda D-$\ew\ fit, 
have \ew\ $<$ 60 m\AA, and have errors less than a third of the \ew.
This set of lines produces a sample of the best weak Fe I lines with 
the most accurate \ew\ and the abundances derived using them will have a 
minimal dependence on the choice of microturbulent velocity.
The second set, subsequently called ``the good line set'', consists of all 
the Fe I lines with errors less than a third of the \ew\ and with \ew\ 
computed from the fit for $\lambda D-$\ew\ determined for each star.
In future papers, the \ew\ of lines of other elements will be determined in the 
same manner as the Fe I lines of ``the good line set''. 
This way a consistent comparison can be done among the resulting abundances
without a restriction on the strength of the lines used.
The ``weak line set'' is different for each star. Actually the weak lines
for the stars near the main sequence are no longer weak for the cooler stars
in our sample.
We compare the results from 20 lines common to 15 stars over almost the whole 
range in \teff\ with the results from the ``weak line set''; there is no 
difference within the errors, nor a trend in \teff. 

For Fe II, the \ew\ of the lines are also determined using the fit to 
the $\lambda D-$\ew\ relation of the Fe I lines of ``the good line set''.
The Fe II lines follow the relationship determined from Fe I lines well,
as shown in Figure 3. 
Additional Fe II lines, not picked up automatically, were measured by hand for
the stars near the turn off and for M71-G53425\_4612 and M71-G53457\_4709. 
The set of \ew\ for the hand selected Fe II lines is computed from the 
$\lambda D-$\ew\ relation of the Fe I lines, after determining their 
observed depth from the spectra.

The number of Fe I lines, for both sets of lines, and the number of Fe II 
lines utilized in this analysis are listed in Table 1.

\subsection{Spectroscopic Effective Temperature}

The effective temperature (\teff) of a star can be determined spectroscopically 
by requiring the abundance to be independent of the lower excitation potential.
This technique can be applied to 20 of our stars where 
we have Fe I lines with enough range in lower excitation potential to do so. 
For the spectroscopic \teff\ determination we are using ``the weak line set'' of Fe I
lines, because its resulting abundance and spectroscopic \teff\ will be 
only weakly dependent on the choice of microturbulent velocity.
We find that the spectroscopic \teff\ is in good agreement 
with the photometric \teff\ derived in Paper I, as shown in Figure 4. 
The solid line in Figure 4 shows the ideal case when the spectroscopic 
and the photometric \teff\ are equal. 
The scatter around the solid line is about 150 K, which is comparable to the
error of the photometric \teff\ of 75 K for giants and of 150 K for dwarfs 
(Paper I), also shown in Figure 4. 
The scatter around the solid line is symmetric, not above or below, 
indicating the lack of systematic effects with the photometric temperatures.

\subsection{Microturbulent Velocity}

The microturbulent velocity (\mtv) of a star can be determined spectroscopically 
by requiring the abundance to be independent of the strength of the lines measured
as the equivalent width. 
We apply this technique for both sets of Fe I lines.
The resulting \mtv\ and the \fe\ computed with it for ``the weak line set'' of 
Fe I lines are listed in Table 1. 
Only 18 of our stars have enough weak Fe I lines to derive \mtv\ 
spectroscopically. 
We also compute \mtv\ using all the good Fe I lines for 20 of our stars.
The difference of the resulting \fe\ computed with the derived \mtv\ for 
the two set of lines  is plotted with respect to \teff\ in Figure 5. 
The mean difference is $-0.004 \pm$ 0.011, hence, as expected, the \fe\ 
results from both sets of lines show a very good agreement.

The relationship between \mtv\ determined for the set of all good Fe I lines 
and the photometric \teff\ is shown in Figure 6. 
The solid line corresponds to a linear least squares fit of the data, 
excluding the red horizontal branch (RHB) stars, marked with circles. 
The best fit line is given by:
$$\xi = 3.30 - 4.66 \times 10^{-4} \times T_{eff} $$
The scatter around the solid line is about 0.2 \kms, which is a reasonable
estimation of the error in \mtv. For the rest of the analysis, we will use
the set of all the good Fe I and Fe II lines, with \mtv\ computed from the 
\mtv-\teff\ fit.
For the RHB stars we use a value of 1.61 \kms, which corresponds to the mean
value determined for the three RHB stars. 
The microturbulent velocity used for our stars is listed in Table 1.

\section{RESULTS}

Given the stellar parameters from Table 1, we determined the iron abundance
using the equivalent widths of all the good Fe I and Fe II lines identified 
in the HIRES spectra. 
We employ the grid of stellar atmospheres from \citet{kur93b} to compute 
the iron abundance using the four stellar atmosphere models with the closest 
\teff\ and \grav\ to each star's parameters.
The \fe\ listed in Table 1 is an interpolation of the results from the 
closest stellar model atmospheres to the appropriate \teff\ and \grav\ for
each star.

\subsection{\fe\ from Fe I lines}

The results of \fe\ from Fe I lines are listed in column 9 of Table 1 and
plotted against the photometric \teff\ in Figure 7a. 
\teff\ is used for the x-axis as a convenient parameter for characterizing the 
position of the stars in the color-magnitude diagram as it also ranks the
stars in luminosity (except for the RHB stars).
The errors listed in Table 1 correspond to the larger of the statistical 
uncertainty, given by the standard deviation of the iron abundance from 
different lines divided by the square root of the number of lines, or a 
minimum value of 0.03 dex.
These errors are lower limits to the actual uncertainties in the abundances,
since they do not include uncertainties due to the stellar parameters nor any 
systematic effects that might be present. 
We estimate the sensitivity of \fe\ derived from Fe I lines with respect to
the stellar parameters in three cases 4250/1.0/1.0, 5000/2.5/1.0 and 5500/4.0/1.0,
where the three numbers correspond to \teff/\grav/\mtv. 
The results are listed in Table 2, where the range adopted for each parameter is
representative of its uncertainty.
Our determination of \fe\ from Fe I lines is most sensitive to errors 
in \teff, which is
less than $\sim$ 0.1 dex for $\Delta$\teff\ of $\pm$ 100 K, and have a
minimal sensitivity on the choice of metallicity of the model atmosphere grid
for plausible changes in \fe\ ($\pm$0.2 dex).

The solid line, shown in Figure 7a, is a linear fit weighted by the errors
of \fe\ versus \teff. 
The slope of the fit is $(-0.8 \pm 3.6) \times 10^{-5}$ dex/K,
which is consistent with \fe\ being constant, independent of \teff\ (ie, of 
luminosity or equivalently position in the color-magnitude diagram).
We divide our sample in four groups of stars: giant stars at or above the
RHB, stars on the RHB, giant stars below the RHB, 
and main sequence stars near the turnoff. The mean \fe\ for each group
is listed on Table 4. We found no significant difference in the mean \fe\
obtained from Fe I lines among the defined groups of stars.

The mean \fe\ weighted by the errors of all 25 stars 
is $-0.71 \pm 0.08$, in very good
agreement with earlier determinations \citep{coh83,gra86,lee87,sne94}.

\subsection{\fe\ from Fe II lines}

The determinations of \fe\ from Fe II lines are listed in column 11 of Table 1 
and plotted against the photometric \teff\ in Figure 7b.
The errors listed in Table 1 corresponds to the statistical uncertainty or
a value of 0.05 dex, whichever is larger.
We estimate the sensitivity of \fe\ derived from Fe II ines with respect to
the stellar parameters in the same manner as the sensitivity of \fe\ from 
Fe I lines. The results are listed in Table 3, where the range adopted for 
each parameter is representative of its uncertainty.
We see a stronger sensitivity on the stellar parameters from the 
Fe II lines than from the Fe I lines.
The \fe\ determination from Fe II lines is most sensitive to the systematic 
error (note that the internal uncertainty in \grav\ is $\leq$ 0.1 dex) in \grav,
as well as to \teff\ among the coolest M71 giants.
The sensitivity on the choice of metallicity of the model atmosphere grid
is small for reasonable changes in metallicity.

The solid line, shown in Figure 7b, is a linear fit weighted by the errors
of \fe\ versus \teff.
The slope of the fit is $(+3.1 \pm 5.2) \times 10^{-5}$ dex/K,
which is consistent with \fe\ being constant, independent of \teff.
We found that there is no significant difference in the mean \fe\ obtained
from Fe II among stars from different luminosity groups, listed in Table 4.
The mean \fe\  weighted by the errors is $-0.84 \pm 0.12$, in very good
agreement with our result from Fe I lines and earlier determinations 
\citep{coh83,gra86,lee87,sne94}.

\subsection{NLTE effects}

The iron abundance could be affected by departures from LTE. The main NLTE
effect in late-type stars is caused by overionization of electron donor
metals by ultraviolet radiation \citep{aum75}.
Recently, \citet{gra99} and \citet{the99} studied NLTE effects in Fe abundances in
metal-poor late-type stars. 
\citet{gra99} found that NLTE corrections for Fe lines are very small in
dwarfs of any \teff, and only small corrections ($<$ 0.1 dex) are expected for 
stars on the red giant branch.
\citet{the99} found that NLTE corrections become less important as [Fe/H]
increases, being less than 0.1 dex for stars with \fe $> -$0.75 dex, and that
ionized lines are not significantly affected by NLTE.

One way to explore possible NLTE effects present in our data is by comparing
the results from Fe I and Fe II lines. The difference between \fe\ from Fe II
and Fe I lines is plotted in Figure 8 against \teff. The solid line is a linear
fit weighted by the errors.
The slope of the fit is $(+2.0 \pm 8.2)\times 10^{-5}$ dex/K, which is
nearly flat. The mean difference is $-0.13 \pm 0.18$. We conclude that NLTE 
effects are negligible in our iron abundance determination, as expected 
from results of earlier studies \citep{gra99,the99}. 

\section{DISCUSSION}

Our \fe\ abundance results involve for the first time a wide luminosity sample 
of stars, which includes at the same time stars from luminous giants to 
stars near the turnoff. 
We find that the \fe\ abundance, from both Fe I and Fe II
lines, is independent of \teff, and equivalently luminosity. 

Our result is in agreement with the work of \citet{gra01}. 
They present abundances from high dispersion spectra from the VLT of 
stars in NGC 6397 and NGC 6752.
They found that the \fe\ obtained for stars at the base of the subgiant
branch agrees within a few percent with the \fe\ obtained for 
stars at the main sequence turnoff, and further compare this 
value with analysis of the RGB stars in this cluster by other groups. 
Note that the luminosity range of
the sample presented in our work is several orders of magnitude wider than
the luminosity range of \citet{gra01}'s sample. 

Our results, and those of \citet{gra01}, appear to be in disagreement with 
inhomogeneities in \fe\ found earlier by \citet{kin98}. 
They obtained \fe=$-$2.52 dex
for a sample of subgiant stars in M92, which is a factor of two smaller than 
\fe\ measurements using red giants in the same cluster \citep{coh79,sne91}.
\citet{kin98} compare their result for the M92 subgiants with analysis
of RBG stars by other groups, who may have
determined the stellar parameters and performed the abundance determinations
in a different way. This possible difference in the analysis of the giant and 
subgiant sample may account for the difference in \fe\ found by \citet{kin98}
or perhaps the determination of the stellar parameters by \citet{kin98} is
flawed. 
Our result, on the other hand, is robust, because we have determined both the 
stellar parameters and the Fe abundance in a homogeneous and consistent manner 
for all our stars.

\citet{gra01} also found that NGC 6397 is homogeneous in both O and Fe, 
while an O-Na anticorrelation is present among unevolved stars 
in NGC 6752, which is very difficult to explain by the deep mixing scenario.
Lines from many additional species, including O, Na, Mg, Ti, Sc among others, 
are observed in our HIRES spectra. 
We plan to present additional information in the matter of
light, iron-peak, and heavy elements in the near future.

\section{CONCLUSIONS}

We present results of a high dispersion analysis of Fe I and Fe II lines to
obtain \fe\ for 25 members of the Galactic globular cluster M71. 
Our sample of stars includes
19 giant stars, 3 horizontal branch stars, and 3 stars near the
turnoff. Our conclusions are summarized as follows:

\begin{itemize}
\item The \fe\ obtained from Fe I lines agrees very well with the 
\fe\ obtained from Fe II lines. 
\item The mean \fe\ obtained from Fe I and Fe II lines of all 25 stars 
is in good agreement with earlier determinations.
\item The \fe\ obtained from both Fe I and Fe II lines is independent 
of \teff, and equivalently luminosity.
\item No difference is found among the mean \fe\ from giant stars 
located at or above the RHB, RHB stars, giant stars located below the RHB and 
stars near the main sequence turnoff.
\end{itemize}

In the near future, we will present the result from the analysis now underway 
of additional elements.

\acknowledgements
The entire Keck/HIRES and LRIS user communities owes a huge debt to 
Jerry Nelson, Gerry Smith, Steve Vogt, Bev Oke, and many other 
people who have worked to make the Keck Telescope and HIRES and LRIS 
a reality and to operate and maintain the Keck Observatory. 
We are grateful to the W. M.  Keck Foundation for the vision to fund
the construction of the W. M. Keck Observatory. 
We thank R. Gratton for providing a detailed description 
of his automatic equivalent width measuring program and R. Pogge
for providing the Gaussian profile fitting routine.
Partial support to MMB was provided by a Theodore Dunham, Jr. grant
for Research in Astronomy and the National Science Foundation under
grant AST-9624680 to MMB and grant AST-9819614 to JGC.

\appendix
\section{Errors in the Equivalent Widths}

The equivalent width of a line is defined as:
$$ W_{\lambda} = \int \frac{g(\lambda)}{cont(\lambda)} d\lambda  
      = \sum_{i} \frac{g(\lambda_{i})}{cont(\lambda_{i})} \Delta \lambda $$
where, $g(\lambda)$ is the flux of the spectrum, $cont(\lambda)$ is the 
continuum level, and $\Delta \lambda $ is the step that can be made 
arbitrarily small to better approximate the integral. 
In this case, $g(\lambda)$ is the Gaussian profile of the line,
given by: 
$$ g(\lambda) = P \times exp \left[ 
\frac{-(\lambda-\lambda_{cen})^{2}}{2\sigma^{2}} \right] $$
where $P$ is the peak of the Gaussian, $\lambda_{cen}$ is the central 
wavelength, and $\sigma$ is the dispersion of the Gaussian. 
If the equivalent width is computed through out the summation then the error
in the equivalent width, $\sigma_{w}$ will be the quadratic summation of the
errors in the individual points times the step:
$$ \sigma_{W_{\lambda}}^{2} = \sum_{i} (\sigma_{i}\times \Delta \lambda)^{2} $$
where:
$$ \sigma_{i}^{2} = 
\left[ \frac{\partial w}{\partial g} \sigma_{gi} \right]^{2}  +
\left[ \frac{\partial w}{\partial cont} \sigma_{cont} \right]^{2} $$
$$ \sigma_{i}^{2} =
\left[ \frac{1}{cont(\lambda_{i})} \sigma_{gi} \right]^{2}  +
\left[ \frac{g(\lambda_{i})}{cont(\lambda_{i})^{2}} \sigma_{cont} \right]^{2} $$
If we assume that the continuum level is equal to one then:
$$ \sigma_{i}^{2} = \sigma_{gi}^{2} + g(\lambda_{i})^{2} \sigma_{cont}^{2} $$
Since the Gaussian depends on three parameters:
$$ \sigma_{gi}^{2} = 
\left[ \frac{\partial  g}{\partial P} \sigma_{P} \right]^{2} +
\left[ \frac{\partial  g}{\partial \lambda_{cen}} \sigma_{\lambda_{cen}} \right]^{2} +
\left[ \frac{\partial  g}{\partial \sigma} \sigma_{\sigma} \right]^{2} $$
$$ \sigma_{gi}^{2} = 
\left[ \frac{g(\lambda_{i})}{P} \sigma_{P} \right]^{2} +
\left[ \frac{(\lambda_{i} - \lambda_{cen})}{\sigma^{2}} g(\lambda_{i}) \ \sigma_{\lambda_{cen}} \right] ^{2} +
\left[ \frac{(\lambda_{i} - \lambda_{cen})^{2}}{\sigma^{3}} g(\lambda_{i}) \ \sigma_{\sigma} \right]^{2} $$
$$ \sigma_{gi}^{2} = g(\lambda_{i})^{2} \times \left[ 
\frac{\sigma_{P}^{2}}{P^{2}} +
\frac{(\lambda_{i} - \lambda_{cen})^{2}}{\sigma^{4}} \sigma_{\lambda_{cen}}^{2} +
\frac{(\lambda_{i} - \lambda_{cen})^{4}}{\sigma^{6}} \sigma_{\sigma}^{2} \right] $$
Finally:
$$\sigma_{i}^{2} = g(\lambda_{i})^{2} \times \left[ 
\frac{\sigma_{P}^{2}}{P^{2}} +
\frac{(\lambda_{i} - \lambda_{cen})^{2}}{\sigma^{4}} \sigma_{\lambda_{cen}}^{2}
+
\frac{(\lambda_{i} - \lambda_{cen})^{4}}{\sigma^{6}} \sigma_{\sigma}^{2} +
\sigma_{cont}^{2} \right] $$

\clearpage

%
%
\begin{deluxetable}{lcccccccccl}
\rotate
\tablenum{1}
\tablewidth{0pt}
\tablecaption{Stellar Parameters for the M71 Sample}
\label{tab1}
\tablehead{
\colhead{ID\tablenotemark{a}} & 
\colhead{\teff} & 
\colhead{\grav} & 
\colhead{\mtv \tablenotemark{b}} &
\colhead{${\rm N_{Fe I}}$ \tablenotemark{b}} &
\colhead{${\rm \fe_{Fe I}}$ \tablenotemark{b}} & 
\colhead{\mtv \tablenotemark{c}} &
\colhead{${\rm N_{Fe I}}$ \tablenotemark{c}} &
\colhead{${\rm \fe_{Fe I}}$ \tablenotemark{c}} &
\colhead{${\rm N_{Fe II}}$ \tablenotemark{d}} &
\colhead{${\rm \fe_{Fe II}}$ \tablenotemark{d}} \\ 
\colhead{} & \colhead{(K)} & \colhead{} & \colhead{(km/s)} & 
\colhead{} & \colhead{}   & \colhead{(km/s)} & \colhead{} & 
\colhead{} & \colhead{} & \colhead{} 
}
\startdata
1--45        & 3950 & 0.90 & 1.48 & 59 &--0.74 $\pm$ 0.03 & 
1.46 & 187 &--0.78 $\pm$ 0.03 & 6 &--0.61 $\pm$ 0.11\\
I            & 4150 & 1.00 & 1.00 & 67 &--0.69 $\pm$ 0.03 & 
1.37 & 186 &--0.76 $\pm$ 0.03 & 6 &--0.88 $\pm$ 0.07\\
1--66        & 4250 & 1.35 & 1.80 & 68 &--0.83 $\pm$ 0.03 & 
1.32 & 179 &--0.71 $\pm$ 0.03 & 6 &--0.86 $\pm$ 0.09\\
1--64        & 4200 & 1.35 & 1.57 & 61 &--0.76 $\pm$ 0.03 & 
1.34 & 187 &--0.74 $\pm$ 0.03 & 5 &--0.73 $\pm$ 0.09\\
1--56        & 4525 & 1.60 & 0.81 & 25 &--0.63 $\pm$ 0.04 & 
1.19 & 127 &--0.56 $\pm$ 0.03 & 2 &--0.83 $\pm$ 0.14\\
1--95        & 4550 & 1.65 & 1.00 & 79 &--0.68 $\pm$ 0.03 & 
1.18 & 184 &--0.67 $\pm$ 0.03 & 8 &--0.87 $\pm$ 0.05\\
1--81        & 4550 & 1.75 & 1.50 & 77 &--0.71 $\pm$ 0.03 & 
1.18 & 180 &--0.64 $\pm$ 0.03 & 6 &--1.05 $\pm$ 0.05\\
1--1         & 4700 & 2.05 & 0.89 & 55 &--0.62 $\pm$ 0.03 & 
1.11 & 134 &--0.67 $\pm$ 0.03 & 5 &--0.86 $\pm$ 0.05\\
1--80\tablenotemark{e,f}        
             & 5300 & 2.45 & ...  & 28 &--0.68 $\pm$ 0.04\tablenotemark{g}& 
1.61 &  71 &--0.69 $\pm$ 0.03 & 5 &--0.92 $\pm$ 0.05\\
1--87\tablenotemark{e}
             & 5300 & 2.45 & 1.62 & 68 &--0.58 $\pm$ 0.03 & 
1.61 & 128 &--0.60 $\pm$ 0.03 & 9 &--0.83 $\pm$ 0.05\\
1--94\tablenotemark{e}
             & 5300 & 2.45 & ...  & 42 &--0.78 $\pm$ 0.03\tablenotemark{g}&
1.61 &  94 &--0.77 $\pm$ 0.03 & 6 &--0.81 $\pm$ 0.05\\
1--60        & 4900 & 2.30 & 0.70 & 49 &--0.77 $\pm$ 0.03 & 
1.02 & 119 &--0.80 $\pm$ 0.03 & 6 &--0.70 $\pm$ 0.05\\
1--59        & 4600 & 2.30 & 1.50 & 62 &--0.79 $\pm$ 0.03 & 
1.16 & 141 &--0.78 $\pm$ 0.03 & 5 &--0.70 $\pm$ 0.05\\
G53476\_4543 & 4900 & 2.65 & 0.83 & 94 &--0.64 $\pm$ 0.03 & 
1.02 & 174 &--0.68 $\pm$ 0.03 & 7 &--0.84 $\pm$ 0.05\\
2--160       & 5100 & 2.70 & 1.10 & 68 &--0.59 $\pm$ 0.03 & 
0.92 & 145 &--0.54 $\pm$ 0.03 & 5 &--0.97 $\pm$ 0.08\\
G53447\_4707 & 5175 & 2.75 & 1.35 & 90 &--0.62 $\pm$ 0.03 & 
0.89 & 155 &--0.57 $\pm$ 0.03 & 7 &--0.86 $\pm$ 0.05\\
G53445\_4647 & 5050 & 2.85 & 0.54 & 50 &--0.54 $\pm$ 0.03 & 
0.95 & 112 &--0.65 $\pm$ 0.03 & 6 &--0.85 $\pm$ 0.05\\
G53447\_4703 & 5000 & 3.00 & 0.90 & 62 &--0.72 $\pm$ 0.03 & 
0.97 & 125 &--0.77 $\pm$ 0.03 & 4 &--0.80 $\pm$ 0.05\\
G53425\_4612 & 5150 & 3.15 & 1.40 & 36 &--0.77 $\pm$ 0.03 & 
0.90 &  80 &--0.73 $\pm$ 0.03 & 2\tablenotemark{h} &--0.91 $\pm$ 0.08\\
G53477\_4539 & 5150 & 3.15 & ...  & 56 &--0.66 $\pm$ 0.03\tablenotemark{g}&
0.90 & 119 &--0.70 $\pm$ 0.03 & 5 &--0.90 $\pm$ 0.05\\
G53457\_4709 & 5200 & 3.35 & 1.24 & 58 &--0.78 $\pm$ 0.03 & 
0.88 &  93 &--0.78 $\pm$ 0.03 & 5\tablenotemark{h} &--0.76 $\pm$ 0.11\\ 
G53391\_4628 & 5100 & 3.35 & ...  & 55 &--0.74 $\pm$ 0.03\tablenotemark{g}&
0.92 & 106 &--0.84 $\pm$ 0.03 & 5 &--0.81 $\pm$ 0.07\\
G53417\_4431 & 5800 & 4.05 & ...  & 19 &--0.66 $\pm$ 0.03\tablenotemark{g}&
0.60 &  38 &--0.68 $\pm$ 0.04 & 3\tablenotemark{h} &--0.61 $\pm$ 0.12\\
G53392\_4624 & 5800 & 4.05 & ...  & 23 &--0.81 $\pm$ 0.04\tablenotemark{g}& 
0.60 &  36 &--0.81 $\pm$ 0.03 & 3\tablenotemark{h} &--0.66 $\pm$ 0.08\\
G53414\_4435 & 5900 & 4.15 & ...  &  5 &--0.82 $\pm$ 0.12\tablenotemark{g}& 
0.55 &  13 &--0.83 $\pm$ 0.04 & 2\tablenotemark{h} &--0.58 $\pm$ 0.17 \\
\enddata
\tablenotetext{a}{Identifications are from \citet{arp71}
or are assigned based on the J2000 coordinates, rh rm rs.s dd dm dd becoming
Grmrss\_dmdd.}
\tablenotetext{b}{Set of weak Fe I lines.}
\tablenotetext{c}{Set of all good Fe I lines.}
\tablenotetext{d}{Set of all good Fe II lines.}
\tablenotetext{e}{RHB star.}
\tablenotetext{f}{Appears to show rotation (Paper I).}
\tablenotetext{g}{Computed with \mtv\ from the set of all good Fe I lines.}
\tablenotetext{h}{Includes additional Fe II lines selected by hand.}
\end{deluxetable}

\clearpage

%
%
\begin{deluxetable}{ccccc}
\tablenum{2}
\tablewidth{0pt}
\tablecaption{Sensitivity of ${\rm \fe_{Fe I}}$ on  Stellar Parameters}
\label{tab2}
\tablehead{
\colhead{} & \colhead{$\Delta$\teff} & \colhead{$\Delta$\grav} & 
\colhead{$\Delta$\mtv} & \colhead{$\Delta$\fe}\\
\colhead{}
& \colhead{+ 100 K} & \colhead{+ 0.2 dex} & 
\colhead{+ 0.2 \kms} & \colhead{+ 0.2 dex} }
\startdata
4250/1.0/1.0\tablenotemark{a} & +0.04 & +0.02 &--0.08\tablenotemark{b} & --0.03 \\
5000/2.5/1.0\tablenotemark{a} & +0.09 & +0.01 &--0.06\tablenotemark{b} & --0.01 \\
5500/4.0/1.0\tablenotemark{a} & +0.08 & +0.02 &--0.03\tablenotemark{b} & --0.01 \\
\enddata
\tablenotetext{a}{\teff/\grav/\mtv}
\tablenotetext{b}{This is for the set of good Fe I lines. 
It is smaller by a factor of 3 for the set of weak Fe I lines.}
\end{deluxetable}

%
%
\begin{deluxetable}{ccccc}
\tablenum{3}
\tablewidth{0pt}
\tablecaption{Sensitivity of ${\rm \fe_{Fe II}}$ on  Stellar Parameters}
\label{tab3}
\tablehead{
\colhead{} & \colhead{$\Delta$\teff} & \colhead{$\Delta$\grav} &
\colhead{$\Delta$\mtv} & \colhead{$\Delta$\fe}\\
\colhead{}
& \colhead{+ 100 K} & \colhead{+ 0.2 dex} &
\colhead{+ 0.2 \kms} & \colhead{+ 0.2 dex} }
\startdata
4250/1.0/1.0\tablenotemark{a} & --0.12 & +0.11 & --0.04 & --0.07 \\
5000/2.5/1.0\tablenotemark{a} & --0.02 & +0.09 & --0.03 & --0.04 \\
5500/4.0/1.0\tablenotemark{a} & --0.03 & +0.08 & --0.02 & --0.03 \\
\enddata
\tablenotetext{a}{\teff/\grav/\mtv}
\end{deluxetable}

\clearpage
%
%
\begin{deluxetable}{ccccc}
\tablenum{4}
\tablewidth{0pt}
\tablecaption{\fe\ for Each Group of Stars}
\label{tab4}
\tablehead{
\colhead{Star group} & \colhead{$<V>$} & \colhead{$N_{stars}$} &
\colhead{$<\fe_{FeI}>$} & \colhead{$<\fe_{FeII}>$} }
\startdata
RGB at or above RHB & 13.46 & 10 & $-0.71\pm$0.07 & $-0.83\pm$0.12 \\
RHB                 & 14.50 &  3 & $-0.68\pm$0.07 & $-0.86\pm$0.05 \\
RGH below RHB       & 15.92 &  9 & $-0.69\pm$0.09 & $-0.85\pm$0.06 \\ 
MS TO               & 17.76 &  3 & $-0.78\pm$0.06 & $-0.64\pm$0.13 \\
\enddata
\end{deluxetable}
                                                 
\clearpage


\begin{figure}
\epsscale{0.7}
\plotone{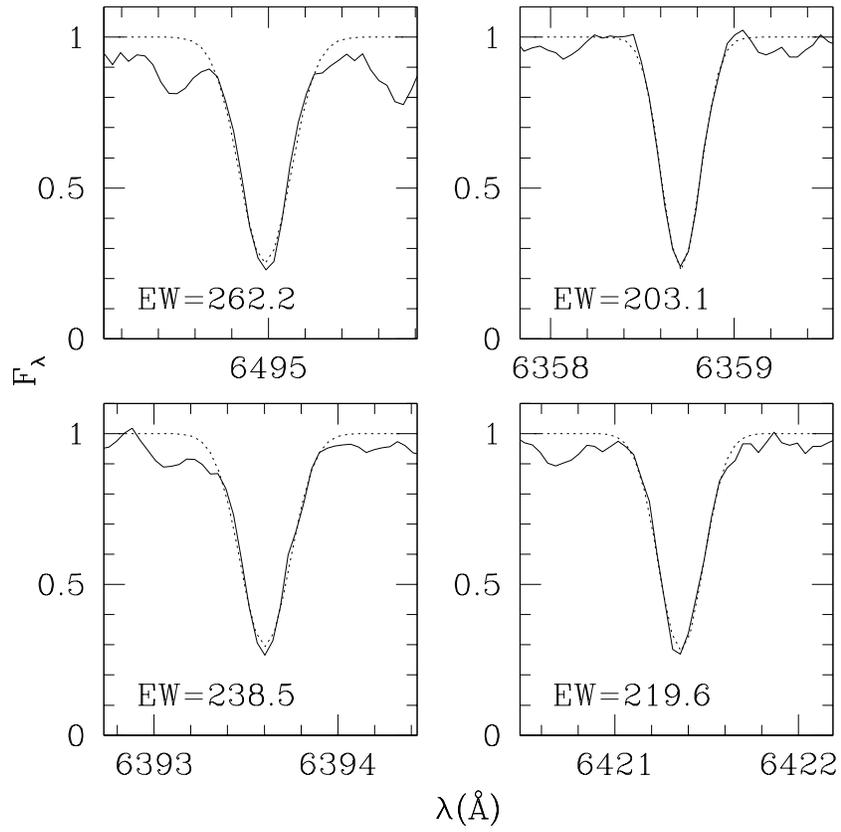}
\caption[fig1.ps]{Strongest observed Fe I lines for M71-1-45. The observed
lines are plotted with a solid line, and the corresponding Gaussian profile
is plotted with a dashed line. The \ew\ of each line is indicated in the 
botton left corner of each pannel.
\label{fig1}}
\end{figure}

\begin{figure}
\epsscale{0.7}
\plotone{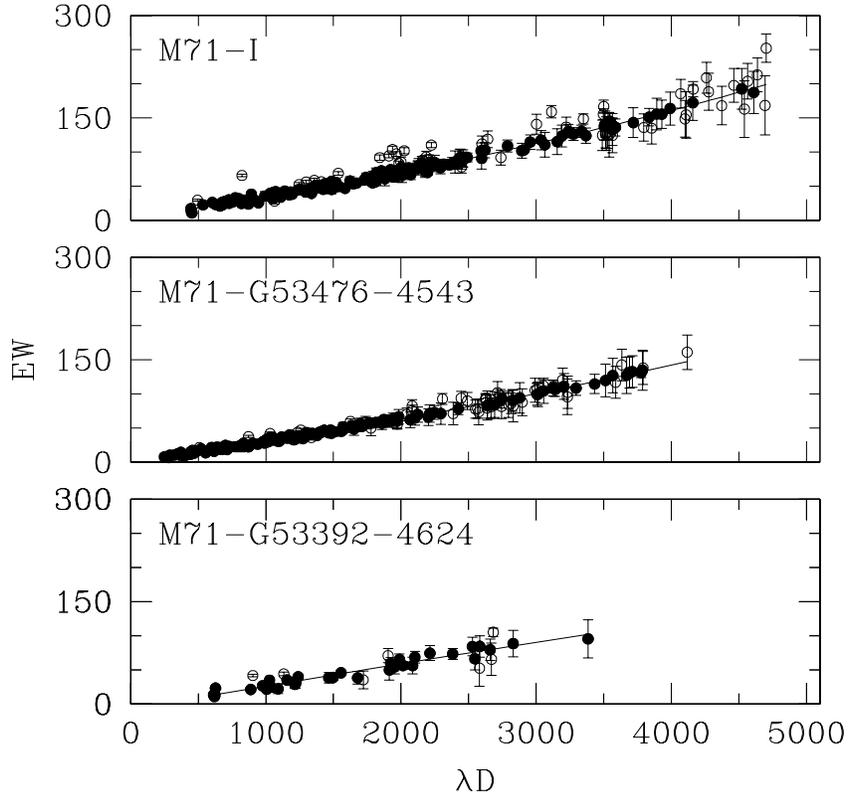}
\caption[fig2.ps]{Central depth times wavelength versus \ew\ relations for
M71-I (one of the most luminous and coolest stars in our sample), 
M71-G53476\_4543 (a star of medium luminosity and temperature), and 
M71-G53392\_4624 (one of the faintest and hottest stars in our sample). 
The solid curve is a second order fit obtained after several 
iterations of rejection of points deviating by 2$\sigma$ or more.
The points used in the fit are shown in black.
\label{fig2}}
\end{figure}

\begin{figure}
\epsscale{0.7}
\plotone{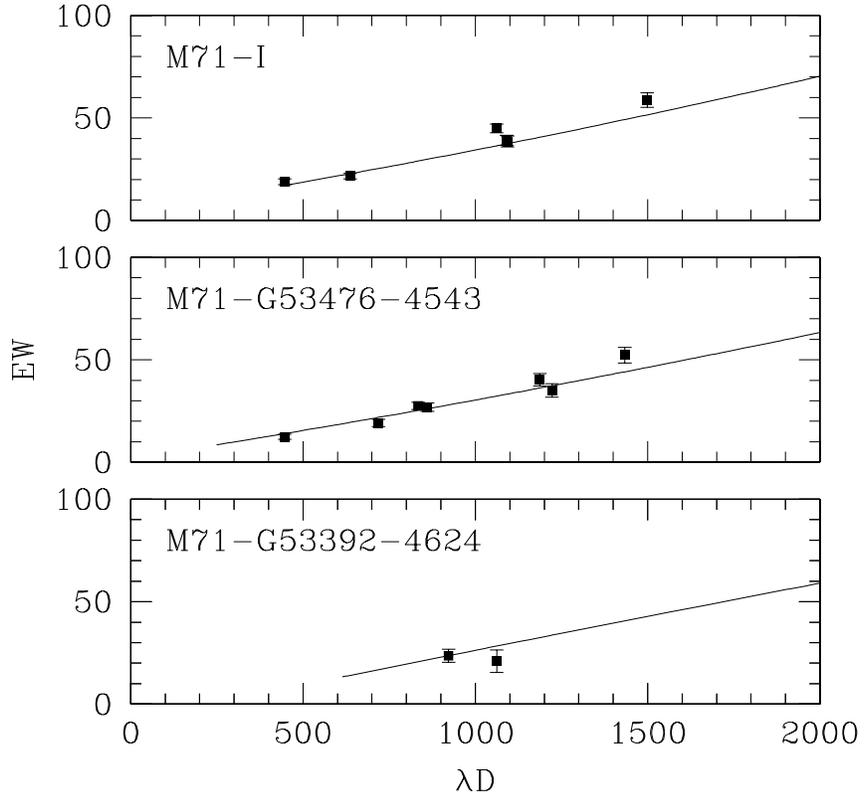}
\caption[fig3.ps]{Central depth times wavelength versus \ew\ relations for
M71-I (one of the most luminous and coolest stars in our sample),
M71-G53476\_4543 (a star of medium luminosity and temperature), and
M71-G53392\_4624 (one of the faintest and hottest stars in our sample).
Solid squares denote identified Fe II lines and 
the solid curve is the second order fit obtained for Fe I lines
and shown in Figure 2.
\label{fig3}}
\end{figure}

\begin{figure}
\epsscale{0.7}
\plotone{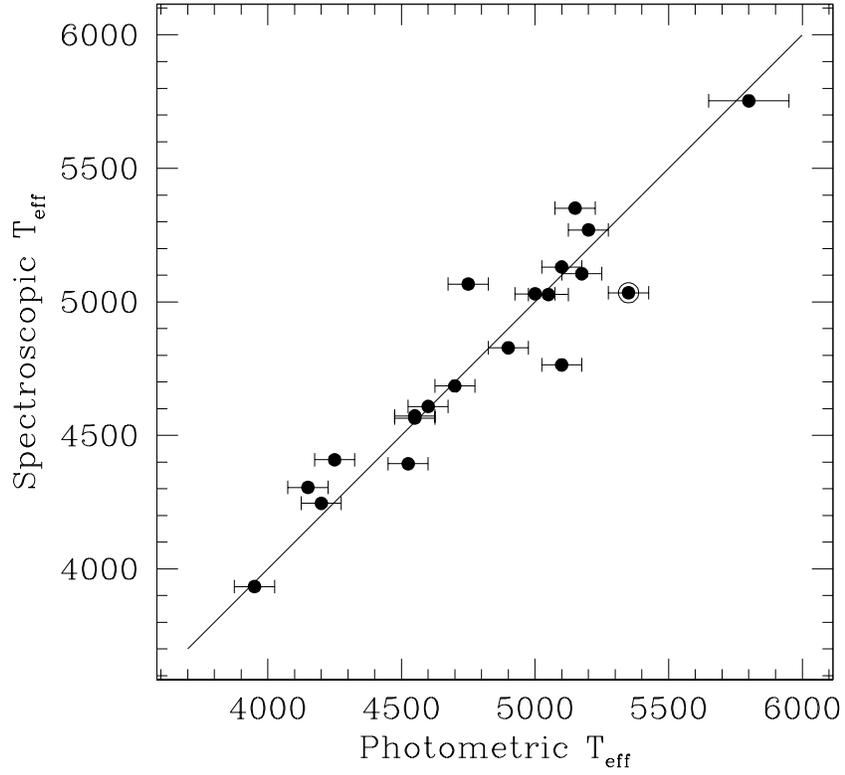}
\caption[fig4.ps]{Photometric \teff\ versus spectroscopic \teff\ for the
M71 sample. The solid line indicates the ideal case when the photometric and 
spectroscopic \teff\ have the same value. The scatter around the solid line is 
about 150 K. The only RHB star for which \teff\ can be determined 
spectroscopically is marked with an open circle.
\label{fig4}}
\end{figure}

\begin{figure}
\epsscale{0.7}
\plotone{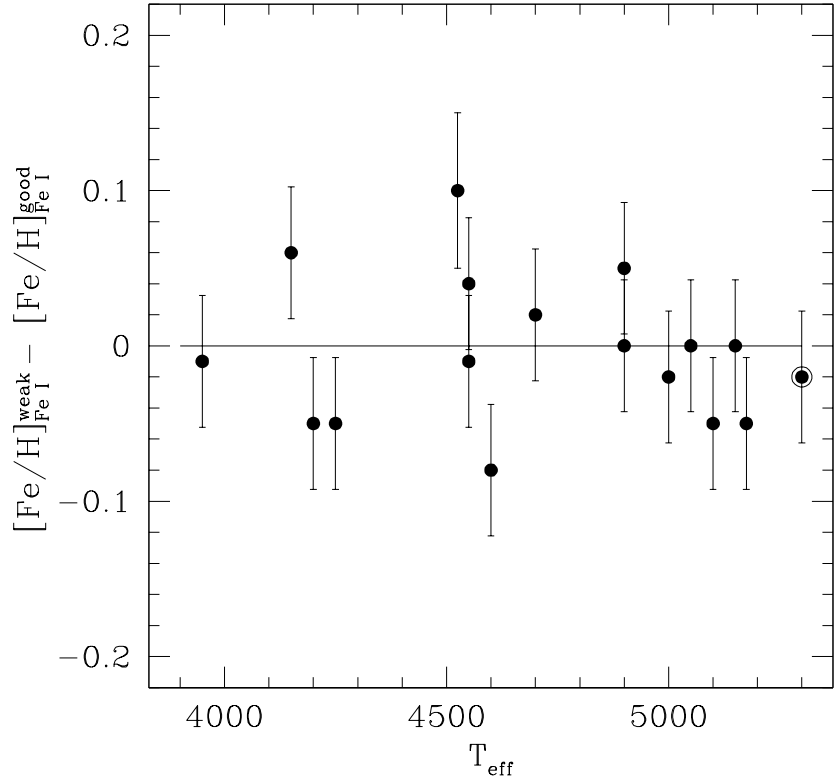}
\caption[fig5.ps]{The difference between \fe\ computed with the derived 
\mtv\ for the set of weak Fe I lines and with \mtv\ from the set of all good 
Fe I lines is plotted with respect to \teff. 
The solid line indicates equality.
The only RHB star for which \mtv\ can be determined spectroscopically
is marked with an open circle.
\label{fig5}}
\end{figure}

\begin{figure}
\epsscale{0.7}
\plotone{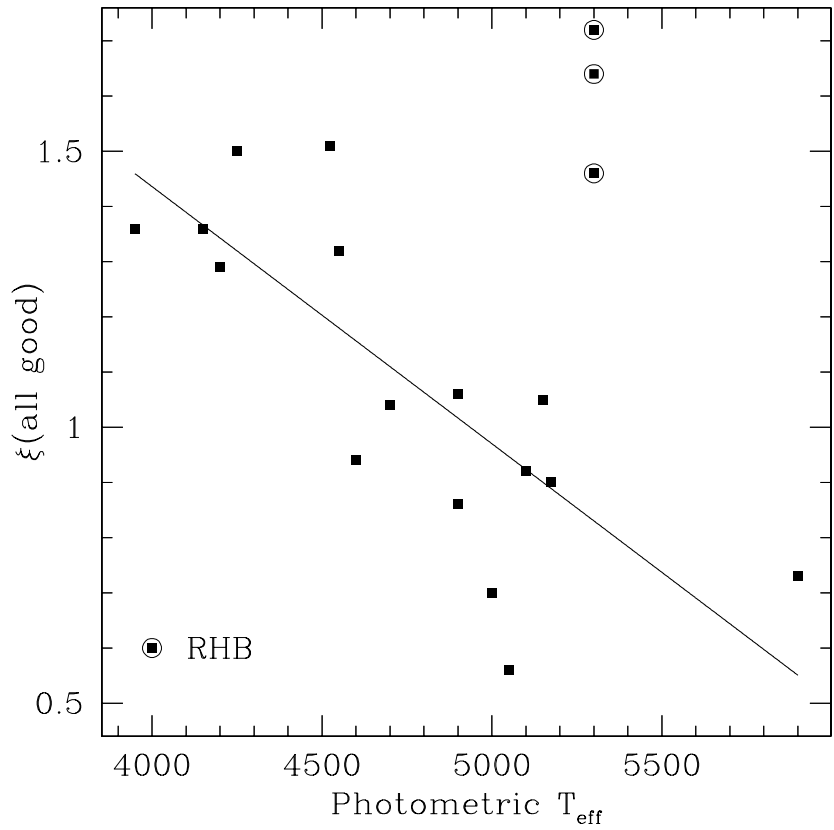}
\caption[fig6.ps]{\mtv\ determined for the set of all good Fe I lines 
is shown as a function of \teff.
The solid line is the linear fit weighted by the errors. 
The RHB stars are excluded from the fit and are marked with circles.
The scatter around the solid line is about 0.2 \kms.
\label{fig6}}
\end{figure}

\begin{figure}
\epsscale{0.7}
\plotone{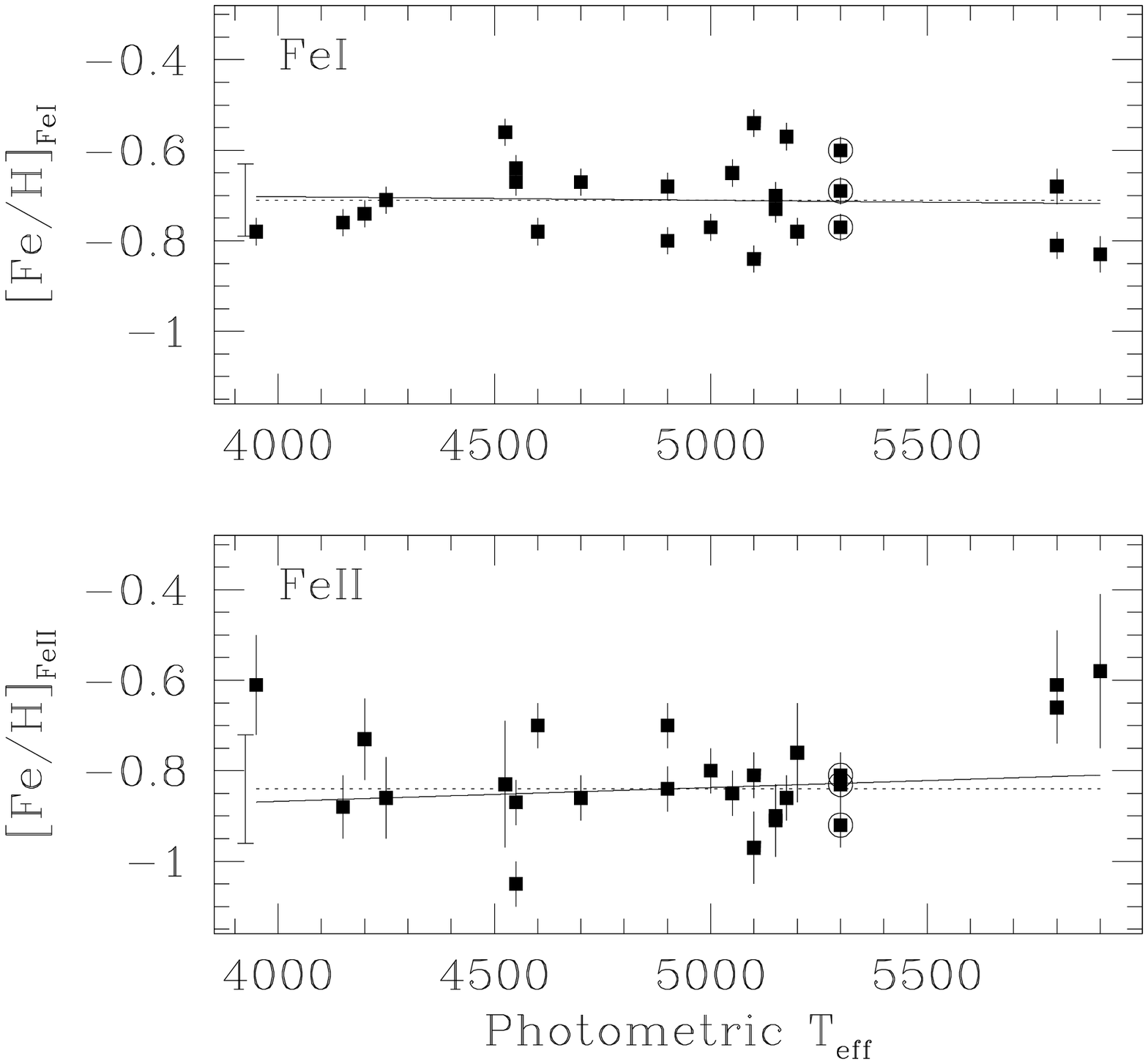}
\caption[fig7.ps]{[Fe/H] from Fe I (upper panel) and Fe II (lower panel) against 
photometric \teff. The solid lines are linear fits weighted by the errors. 
In both cases, \fe\ shows no dependence with \teff. The dashed lines indicate the
mean \fe\ with their respective error plotted as an error bar at 3925 K.
Note that $<\fe_{Fe I}> = -0.71 \pm 0.08$ and $<\fe_{Fe II}> = -0.84 \pm 0.12$.
The RHB stars are marked with a open circles.
\label{fig7}}
\end{figure}

\begin{figure}
\epsscale{0.7}
\plotone{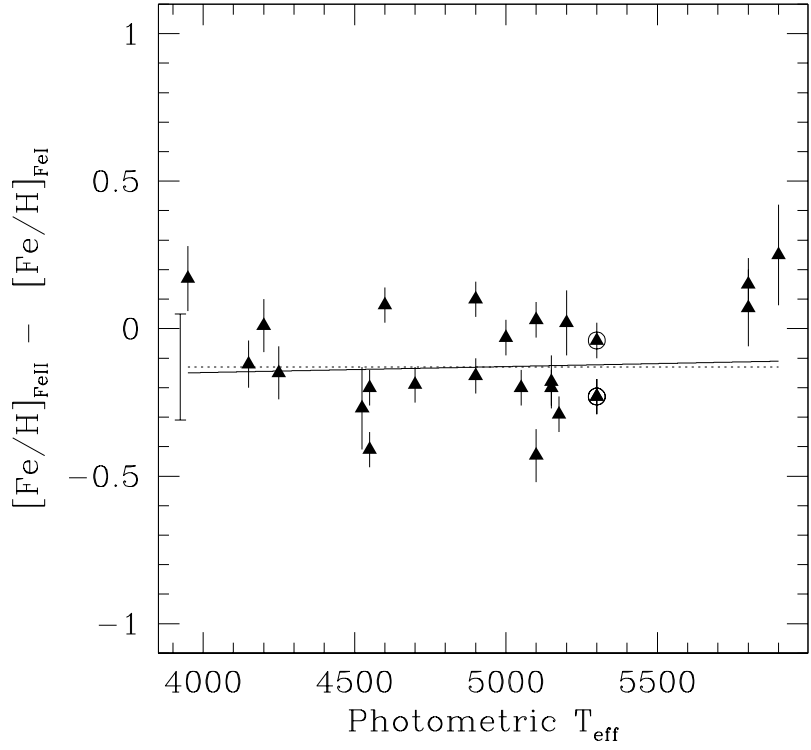}
\caption[fig8.ps]{Difference between [Fe/H] from Fe I and Fe II against \teff.
The solid line, which is nearly flat, is a linear fit weighted by the errors.
The dashed line indicates the mean difference with its respective error plotted 
as an error bar at 3925 K. Note that the mean difference is $-0.13 \pm 0.18$. 
The RHB stars are marked with a open circles.
\label{fig8}}
\end{figure}

\end{document}